\documentclass[conference]{IEEEtran}
\IEEEoverridecommandlockouts

\usepackage{cite}
\usepackage[dvipsnames]{xcolor}

\usepackage{amsmath,amssymb,amsfonts}
\usepackage{algorithmic}
\usepackage{lipsum,adjustbox}
\usepackage{verbatim}
\usepackage{graphics}
\usepackage{nicefrac}
\usepackage{stackengine}
\usepackage{pgfplots}
\pgfplotsset{width=7cm,compat=1.8}
\usepackage{multirow}
\usepackage{soul}
\usepackage{amssymb}
\usepackage{amsmath}
\usepackage{siunitx}

\usepackage{algorithm}
\usepackage{algorithmic}
\usepackage{stmaryrd}
\usepackage{graphicx}
\usepackage{mathtools} 
\usepackage{textcomp}
\usepackage{booktabs}
\usepackage{stfloats}
\usepackage{scalefnt}
\usepackage{mathrsfs}
\def\BibTeX{{\rm B\kern-.05em{\sc i\kern-.025em b}\kern-.08em
    T\kern-.1667em\lower.7ex\hbox{E}\kern-.125emX}}

\usepackage[nolist]{acronym}
\begin{acronym}

\acro{BER}{bit error rate}
\acro{QoS}{quality of service}
\acro{UAV}{unmanned aerial vehicle}
\acro{IoT}{internet of things}
\acro{FBS}{flying base station}
\acro{MARL}{multi-agent reinforcement learning}
\acro{ML}{machine learning}
\acro{AI}{artificial intelligence}
\acro{RL}{reinforcement learning}
\acro{UE}{user equipment}
\acro{SNR}{signal-to-noise ratio}
\acro{SINR}{signal-interference-to-noise ratio}
\acro{DQN}{deep Q-network}
\acro{PO-MDP}{partially observable Markovian decision process}
\acro{QoE}{quality of experience}
\acro{EC-MARL}{emerging communications in multi-agent reinforcement learning}
\acro{LoS}{line of sight}
\acro{NLoS}{non-line-of-sight}
\acro{LSTM}{long short-term memory}
\acro{FPR}{flying passive radar}
\acro{BS}{base station}
\acro{ISAC}{integrated sensing and communication}
\acro{OFDM}{orthogonal frequency division multiplexing}
\acro{RCS}{radar cross-section}
\acro{NN}{neural network}
\acro{LSTM}{long short-term memory}
\acro{MLP}{multi-layer perceptron}
\acro{FFN}{feed-forward network}
\acro{ICI}{inter-carrier interference}
\acro{CTDE}{centralized training with decentralized execution}
\end{acronym}

 \usepackage{bm}

\usepackage[utf8]{inputenc}
\usepackage[english]{babel}
\usepackage{amsthm}
\usepackage{commath}
\usepackage{subcaption}
\usepackage[left=1.62cm,right=1.62cm,top=0.75in, bottom=1.1in]{geometry}
\usepackage{bm}

\usepackage{accents}
\newcommand{\dbtilde}[1]{\accentset{\approx}{#1}}

\begin{document}

\title{
Reinforcement Learning for Enhancing Sensing Estimation in Bistatic ISAC Systems with UAV Swarms
\thanks{
Obed Morrison Atsu is with the Engineering Division, New York University (NYU) Abu Dhabi, UAE.
Salmane Naoumi is with NYU Tandon School of Engineering, Brooklyn, 11201, NY.
Roberto Bomfin is with the Engineering Division, New York University (NYU) Abu Dhabi, UAE.
Marwa Chafii is with the Engineering Division, New York University (NYU) Abu Dhabi, UAE and NYU WIRELESS, NYU Tandon School of Engineering, Brooklyn, NY. 
This work is supported in part by the NYUAD Center for Artificial Intelligence and Robotics, funded by Tamkeen under the Research Institute Award CG010.
Authors acknowledge the Technology Innovation Institute (TII) for funding this project.
}
}

\author{Obed Morrison Atsu, Salmane Naoumi, Roberto Bomfin, Marwa Chafii}  
\maketitle
\begin{abstract}
This paper introduces a novel Multi-Agent Reinforcement Learning (MARL) framework to enhance integrated sensing and communication (ISAC) networks using unmanned aerial vehicle (UAV) swarms as sensing radars. By framing the positioning and trajectory optimization of UAVs as a Partially Observable Markov Decision Process, we develop a MARL approach that leverages centralized training with decentralized execution to maximize the overall sensing performance. Specifically, we implement a decentralized cooperative MARL strategy to enable UAVs to develop effective communication protocols, therefore enhancing their environmental awareness and operational efficiency. Additionally, we augment the MARL solution with a transmission power adaptation technique to mitigate interference between the communicating drones and optimize the communication protocol efficiency. Moreover, a transmission power adaptation technique is incorporated to mitigate interference and optimize the learned communication protocol efficiency. Despite the increased complexity, our solution demonstrates robust performance and adaptability across various scenarios, providing a scalable and cost-effective enhancement for future ISAC networks.
\end{abstract}

\section{Introduction} 
\label{sec:introduction}
Within the emerging landscape of 6G systems, \ac{ISAC} has been identified as a key technology shaping the future of wireless systems. \ac{ISAC} represents an innovative framework that enables the development of spectrally coexistent communication and sensing functionalities, thereby enhancing spectrum efficiency and reducing hardware and computational resource costs \cite{LiuJSAC}.
Communication-centric \ac{ISAC} is a major area of research, aiming to add opportunistic sensing capabilities to existing communication infrastructures. This approach focuses on meeting communication application requirements while estimating sensing parameters from communication waveforms reflected off objects in the environment, effectively repurposing the channel estimation process \cite{zhang2022accelerating}.
Pertinent to communication-centric \ac{ISAC} networks is the use of sensing receivers in two primary configurations, either monostatic or decentralized. Decentralized configurations, such as bistatic or multi-static setups, involve passive radars that are separately placed with the aim of localizing and tracking targets in the environment without disrupting the existing communication infrastructure. These configurations have shown significant promise in practical settings compared to monostatic systems, which often experience high self-interference leakage \cite{jump}. However, several challenges, such as asynchrony, limited visibility, high double path loss, and blind spots, hinder the practical application of \ac{ISAC} in decentralized settings. Therefore, integrating non-terrestrial components, such as \acp{UAV}, into terrestrial \ac{ISAC} networks is a promising approach to overcome these issues.

Recently, there has been growing interest from both industry and academia in integrating \acp{UAV} into future mobile infrastructure due to their potential to revolutionize networks and enhance a wide range of applications, from emerging 6G use cases to critical military and public safety operations \cite{surveyUAVs}. 
Moreover, \acp{UAV} offer dynamic reconfigurability, adaptability, cost-effectiveness, and rapid deployment, making them ideal for communication-centric \ac{ISAC} applications, including real-time monitoring and public safety operations such as search-and-rescue missions \cite{JAndrewZhang2020EnablingJC}.
Several studies have explored the use of \acp{UAV} to enhance wireless networks. For example, \cite{uav_bs1} proposed deploying a \ac{UAV} to provide wireless coverage for indoor users in high-rise buildings during disaster situations. \cite{yang2022path} investigated trajectory planning for rescue relief tasks and introduced a deep reinforcement learning algorithm based on intrinsic rewards to maximize communication coverage for mobile users. In the specific context of \ac{ISAC}, \cite{isac_uav1} proposed an extended Kalman filtering-based tracking scheme for a \ac{UAV}-enabled \ac{ISAC} system, where a \ac{UAV} tracks a moving object while also communicating with a device attached to it. The work in \cite{Naoumi2023TANAGERS} introduced a \ac{MARL} framework that leverages emergent communication strategies for the effective deployment of \acp{UAV} in \ac{ISAC} settings. Additionally, \cite{isac_uav2} proposed a framework for \ac{UAV}-assisted sensing of ground targets, discussing the \ac{ISAC} interactions between \acp{UAV} and \acp{BS}.
To the best of our knowledge, no existing work has explored the deployment of a collaborative \ac{UAV} swarm for \ac{ISAC} purposes while considering imperfect communication between \acp{UAV} and realistic wireless channels in the swarm's communication.

Inspired by the aforementioned challenges in \ac{ISAC} systems, this work proposes a novel \ac{MARL}-based framework to optimize the deployment of \acp{UAV} for sensing tasks. The framework strategically optimizes \ac{UAV} positioning to enhance radar sensing metrics, such as the total achieved sensing \ac{SNR}. The key contributions of this work are summarized as follows:
\begin{itemize}
    \item We formulate the problem of \ac{UAV} positioning and path planning as a \ac{PO-MDP}.
    \item We introduce a decentralized cooperative \ac{MARL} approach, enabling \acp{UAV} to learn efficient and robust communication protocols, accounting for realistic wireless conditions.
    \item We enhance the \ac{MARL} algorithm with a transmission power adaptation mechanism to mitigate \ac{ICI} between \acp{UAV} and maximize the \ac{SINR}.
\end{itemize}
The subsequent sections of the paper are organized as follows. In Section~\ref{sec:sys_model}, we introduce the system model for both the sensing \ac{ISAC} network and the communication network between \acp{UAV}. Section~\ref{sec:proposed_algorithm} formulates the \ac{PO-MDP} for the optimization problem and presents the \ac{MARL} algorithm augmented with learned communication. A comprehensive performance analysis is then conducted in Section~\ref{sec:numerical_results}. Finally, Section~\ref{sec:conclusions} provides concluding remarks and insights.

\section{System Model}
\label{sec:sys_model}
In this study, we consider a mobile network with $K$ \acp{BS} serving users in a predefined area and monitoring $q$ targets. We employ a \ac{UAV} swarm-enabled \ac{ISAC} system for broader sensing coverage. In this setup, $M$ \acp{UAV} are deployed in a decentralized manner as sensing radars. These \acp{UAV} leverage downlink \ac{OFDM} symbols from the \acp{BS} and repurpose the channel estimation process to estimate the sensing parameters of objects in the environment.

\subsection{Targets sensing channel}
From the sensing perspective, the proper trajectory design of each drone in the swarm is the primary factor to enhance the overall sensing performance and therefore optimally covering the designated geographical area. Indeed, at a given time snapshot \( t \), where \acp{UAV} are located at \( \{\pmb{p}_{m}^{t}\}_{m=1}^{M} \), the overall sensing performance of the \ac{UAV} swarm can be measured in terms of the total achieved sensing \ac{SNR}, expressed as
\begin{equation}
\label{eq:total_snr}
\begin{split}
	 \pmb{\mathcal{R}}^{t}
        =
	\sum\limits_{m=1}^{M}
	\sum\limits_{k=1}^{K}
	\sum\limits_{i=1}^{q}
	\pmb{\gamma}_{m,k}^{t} \left( i \right)
	\Pi_{m,k}^{t}(i),
\end{split}
\end{equation}
where \(\pmb{\gamma}_{m,k}^{t}(i)\) is the sensing \ac{SNR} of the \( i^{\text{th}} \) target by the \( m^{\text{th}} \) \ac{UAV} exploiting communication signals transmitted by the $k^{th}$ \ac{BS}, computed as
\begin{equation*}
\label{eq:snr_uac}
\begin{split}
  \pmb{\gamma}_{m,k}^{t}  \left( i \right)
    & = 
   {\mathcal{P}}_{tx}^{k}
   +
   {\mathcal{G}}_{tx}^{k}
   +
   {\mathcal{G}}_r^{m}
   -
   {\mathcal{P}}_{n}
   + 10 \log_{10} \left(\frac{c^{2}}{\left(4 \pi\right)^{3} f_{c}^2 } \right) 
   \\ & 
   + 
   {\mathcal{\sigma}_{i}}
   -  
   20 \log_{10} \left(d_{k,i}d_{i,m}\right),
\end{split}
\end{equation*}
\noindent where ${\mathcal{P}}_{tx}^{k}$ and ${\mathcal{G}}_{rx}^{m}$  represent the transmission power and antenna gain at the the $k^{th}$ \ac{BS}, respectively, ${\mathcal{G}}_r^{m}$ is the antenna gain at the $m^{th}$ \ac{UAV}, ${\mathcal{P}}_{n}$ is the average noise power, and \(c\) and \(f_{c}\) are the speed of light and carrier frequency, respectively. The \ac{SNR} is influenced by the \ac{RCS} of the target, denoted by ${\mathcal{\sigma}_{i}}$, modeled as a statistical distribution and measured in dBsm. The distance from the \ac{BS} to the target, denoted as \(d_{k,i}\), and the distance from the target to the \ac{UAV}, denoted as \(d_{i,m}\), are both measured in meters. Moreover, $\Pi_{m,k}^{t}(i)$ is a binary function indicating whether the \(i^{\text{th}} \) target can be sensed by the \ac{UAV}, provided that the resulting \ac{SNR} is greater than a threshold \(\pmb{\gamma}_{s}\). This threshold determines the \ac{SNR} value above which \acp{UAV} can accurately estimate the target sensing parameters. Importantly, each \ac{UAV} can resolve a specific maximum number of targets at each time step, as their sensing resolution in both delay and angle dimensions is strongly related to the bandwidth and array apertures at both the \acp{BS} and \acp{UAV}. We denote by $q_{\max}$ the maximum number of targets that each \ac{UAV} can sense at each time step of the simulation, and accordingly 
\begin{equation}
 \Pi_{m,k}^{t}(i) = \begin{cases} 1 \quad \text{ if $\pmb{\gamma}_{m,k}^{t}(i) \geq \pmb{\gamma}_{s} $, $ \lfloor \pmb{\gamma}_{m,k}^{t}(i) \rfloor \leq q_{\max} $} \\ 0 \quad \text{ otherwise},\end{cases} \nonumber 
\end{equation}
where $\lfloor \pmb{\gamma}_{m,k}^{t}(i) \rfloor$ represents the sorted index of the sensing \ac{SNR} of the link between the $i^{th}$ target and the $k^{th}$ \ac{BS}, compared to the \ac{SNR} of all other targets from all other links at the $m^{th}$ \ac{UAV} at time $t$.
In the simulation, \acp{UAV} are deployed for up to \(T_{\max}\) time steps. They must collaborate to plan paths and estimate sensing parameters for randomly distributed targets, aiming to maximize the total achieved sensing \ac{SNR}. Therefore, robust inter-communication among the \acp{UAV} is essential for optimal path planning, particularly in a decentralized setup.
\subsection{\acp{UAV} communication channels}
In our scenario, \acp{UAV} operate based on their local observations without a central control unit. Therefore, we assume that the \acp{UAV} have the capability to communicate. In fact, inter-\ac{UAV} communication is crucial for fostering cooperation and sharing environmental knowledge. Effective communication mitigates the limitations of partial observability and enhances sensing accuracy by disseminating information about the environment's topology.
Our approach enables \acp{UAV} to develop message policies throughout their interactions with the environment, rather than relying on predefined communication protocols. This emergent communication strategy allows \acp{UAV} to create their own communication semantics, which are essential for learning effective path planning strategies \cite{chafii2023emergent}.
Indeed, the \acp{UAV} learn to send continuous communication messages, represented as $m_i \in \mathbb{R}^{s}$, where $s$ is the dimension of the communication protocol. These messages, generated by \acp{NN}, contain comprehensive information about the current state of the environment and the semantics developed by each \ac{UAV}, ultimately enhancing the quality of path planning.
Additionally, unlike previous studies that assume perfect wireless channels between all \ac{UAV} pairs, our research focuses on practical applications with realistic models of \ac{UAV} communication channels. The transmission process is stochastic and influenced by factors such as network conditions, including path loss, limited shared bandwidth, and interference, as well as the mobility of the \acp{UAV}. As a result, messages sent by one \ac{UAV} may not be received by all others. Therefore, the \ac{UAV} swarm must learn effective and robust path planning strategies that address the complexities of realistic wireless communication environments.

For the \acp{UAV} communication network, we assume that it accommodates all the \acp{UAV}, with the total dedicated bandwidth equally divided into $M$ equal-sized overlapping channels. The $m^{th}$ \ac{UAV} operates on channel $\mathcal{C}_{m}$, defined by the carrier frequency  $f_{\widetilde{m}}$, for the entire flight duration. Furthermore, we consider a non-orthogonal channel partition, which results in \ac{ICI} between channels in addition to the path loss experienced by the \acp{UAV}. Inspired by the settings in both \cite{comm_pl} and \cite{SINRpaper}, considering a pair of \acp{UAV}, $m$ and $j$, where the $m^{th}$ \ac{UAV} is transmitting a message at time step $t$, the received \ac{SINR} $\pmb{\Gamma}_{m,j}^{t}$ at the $j^{th}$ \ac{UAV} operating on channel $\mathcal{C}_j$ is given by
\begin{equation}
\label{eq:sinr_uavs}
\begin{split}
  \pmb{\Gamma}_{m,j}^{t} & = 
   {\mathcal{P}}_{r}^{j,m,t}
   - {\mathcal{I}_{j,\text{-}m}^{t}}
   - {\mathcal{P}}_{n}
   - 30,
\end{split}
\end{equation}
\noindent where ${\mathcal{P}}_{n}$ is the average noise power and ${\mathcal{P}}_{r}^{j,m,t}$ is the signal power at the $j^{th}$ \ac{UAV} transmitted by the $m^{th}$ \ac{UAV} and is computed as
$
{\mathcal{P}}_{r}^{j,m,t} = {\mathcal{P}}_{t}^{m,t} - {\mathcal{P}\mathcal{L}}_{m,j}^{t},
$
\noindent where ${\mathcal{P}}_{t}^{m,t}$ is the transmit power of the $m^{th}$ \ac{UAV} and ${\mathcal{P}\mathcal{L}}_{m,j}^{t}$ is the path loss between the pair of \acp{UAV} at positions $p_m^{t} = \left(x_{m}^{t}, y_{m}^{t}, z_{m}^{t} \right)$ and $p_j^{t} = \left(x_{j}^{t}, y_{j}^{t}, z_{j}^{t} \right)$, computed as
\begin{equation*}
  {\mathcal{P}\mathcal{L}}_{m,j}^{t} = 68.08 + 22.5 \log_{10}(\lVert p_m^{t} - p_j^{t} \rVert) + \xi_{\sigma},  
\label{eq::comm_pl}
\end{equation*}
\noindent where $\xi_{\sigma}$ is a shadow fading term following a Gaussian distribution with zero mean and a standard deviation $\sigma$.
Furthermore, ${\mathcal{I}_{j,\text{-}m}^{t}}$ is the interference power from all other \acp{UAV} in the network on channel $\mathbf{C}_j$, given by
\begin{equation}
\label{eq:interference}
	 {\mathcal{I}_{j,\text{-}m}^{t}} = \sum_{\mathclap{\substack{l=1 \\ l \neq m}}}^{M} {\mathcal{P}}_{r}^{j,l,t} - {\mathcal{T}} \left(\mathcal{C}_{j}, \mathcal{C}_{l} \right),
\end{equation}
where ${\mathcal{T}} \left(\mathcal{C}_{j}, \mathcal{C}_{l} \right)$ is the power attenuation on channel $\mathcal{C}_j$ with respect to the $l^{th}$ \ac{UAV}, depending on the spectral distance between the two channels, as detailed in Table \ref{table:attenuation}.
\begin{table}[!t]
    \centering
    \caption{Attenuation as a function of the spectral distance between two channels $\mathcal{C}_j$ and $\mathcal{C}_l$ with carrier indices $\widetilde{j}$ and $\widetilde{l}$, respectively.}
    \renewcommand{\arraystretch}{1.}
    \begin{tabular}{>{\centering\arraybackslash}m{5.2cm} >{\centering\arraybackslash}m{2.5cm}}
        \toprule
        \textbf{Spectral Distance $|\widetilde{j} - \widetilde{l}|$} & \textbf{Attenuation (dB)} \\
        \midrule
        0 & 0 \\
        1 & 20 \\
        2 & 40 \\
        3 & 50 \\
        4 & 60 \\
        $\geq 5$ and $\left|\frac{f_{\widetilde{j}} - f_{\widetilde{l}}}{f_{\widetilde{j}}}\right| \leq 0.05$ & 95 \\
        else & 110 \\
        \bottomrule
    \end{tabular}
    \label{table:attenuation}
\end{table} 
Furthermore, we define two thresholds, $\pmb{\Gamma}_{1} < \pmb{\Gamma}_{2}$, to evaluate the \ac{SINR} $\pmb{\Gamma}_{m,j}^{t}$ between two agents, $m$ and $j$, during each communication round. Specifically, if $\pmb{\Gamma}_{m,j}^{t} \geq \pmb{\Gamma}_{2}$, the message is successfully received and decoded by agent $j$. If $\pmb{\Gamma}_{1} < \pmb{\Gamma}_{m,j}^{t} < \pmb{\Gamma}_{2}$, the receiver detects that a message was sent but cannot decode its content. Conversely, if $\pmb{\Gamma}_{m,j}^{t} < \pmb{\Gamma}_{1}$, the receiver neither recognizes that a message was transmitted nor decodes it.
\section{Proposed MARL Algorithm} 
\subsection{\ac{PO-MDP} formulation}
\label{subsec:pomdp_formulation}
The objective of this proposed framework, as outlined in Section~\ref{sec:sys_model}, is to maximize the sensing \ac{SNR} of a set of targets within an environment by optimizing the path planning of a \ac{UAV} swarm consisting of $M$ \acp{UAV}, exploiting communication signals from $K$ \acp{BS} serving users in the downlink. To address this problem, we introduce an \ac{MARL} framework where the \acp{UAV}, i.e, \ac{RL} agents, collaboratively learn optimal action policies, specifically path planning strategies, within the simulation environment. Indeed, we mathematically model the environment as a multi-agent finite-horizon decentralized \ac{PO-MDP} with communication, defined by the tuple $\left( \mathcal{S}, \mathcal{A}, \mathcal{O}, T, \mathcal{M}, R, T_{\max}, \gamma \right)$.
Here, $\mathcal{S}$ denotes the space of global states of the simulation environment, and $\mathcal{O} = \left( \mathcal{O}_{1}, \mathcal{O}_{2}, \ldots, \mathcal{O}_{M} \right)$ represents the space of partial observations for each of the \acp{UAV}. The action space is defined as $\mathcal{A} = \left( \mathcal{A}_{1}, \mathcal{A}_{2}, \ldots, \mathcal{A}_{M} \right)$. The stochastic global state transition model is denoted by $T: \mathcal{S} \times \mathcal{A}_{1} \times \mathcal{A}_{2} \times \ldots \times \mathcal{A}_{M} \rightarrow \mathcal{S}$, and $\mathcal{M}$ represents the communication space.
At each time step $t$, the $m^{th}$ \ac{UAV} takes an action $a_{m}^{t}$ based on its local observation $o_{m}^{t}$ and observes the total reward $\pmb{\mathcal{R}}^{t}$ as defined in Eq.~\eqref{eq:total_snr}. 
Although actions are taken separately by each agent, all \acp{UAV} receive the same collective reward $\pmb{\mathcal{R}}^{t}$, which is shared equally across the team, therefore encouraging a cooperative behavior.
The \ac{UAV} aims to learn a communication-based policy $\pi_{\left(\mathcal{A}^{m}, \mathcal{M}^{m}\right)}^{\Theta} \left( \cdot \mid o_{m}^{t} \right)$, parameterized by learnable parameters $\Theta$, to maximize the expected total cumulative discounted reward corresponding to the overall sum of the targets \ac{SNR}, determined by the flight time duration $T_{\max}$ and a discount factor $\gamma \in [0, 1]$, and computed as
\begin{equation}
\label{eq:discounted_reward}
\begin{split}
	 J(\Theta)
        =
    \mathbb{E}_{\pi_{\left({\mathcal{A}},{\mathcal{M}}\right)}^{\Theta}}
    \left[ 
	\sum\limits_{t=1}^{T_{\max}}
    \gamma^{t-1}
    \pmb{\mathcal{R}}^{t}
    \right].
\end{split}
\end{equation}
\subsection{\ac{MARL} architecture}
\label{subsec:marl_architecire}
\label{sec:proposed_algorithm}
\begin{figure*}[t]
    \centering
    \includegraphics[width=0.8\linewidth]{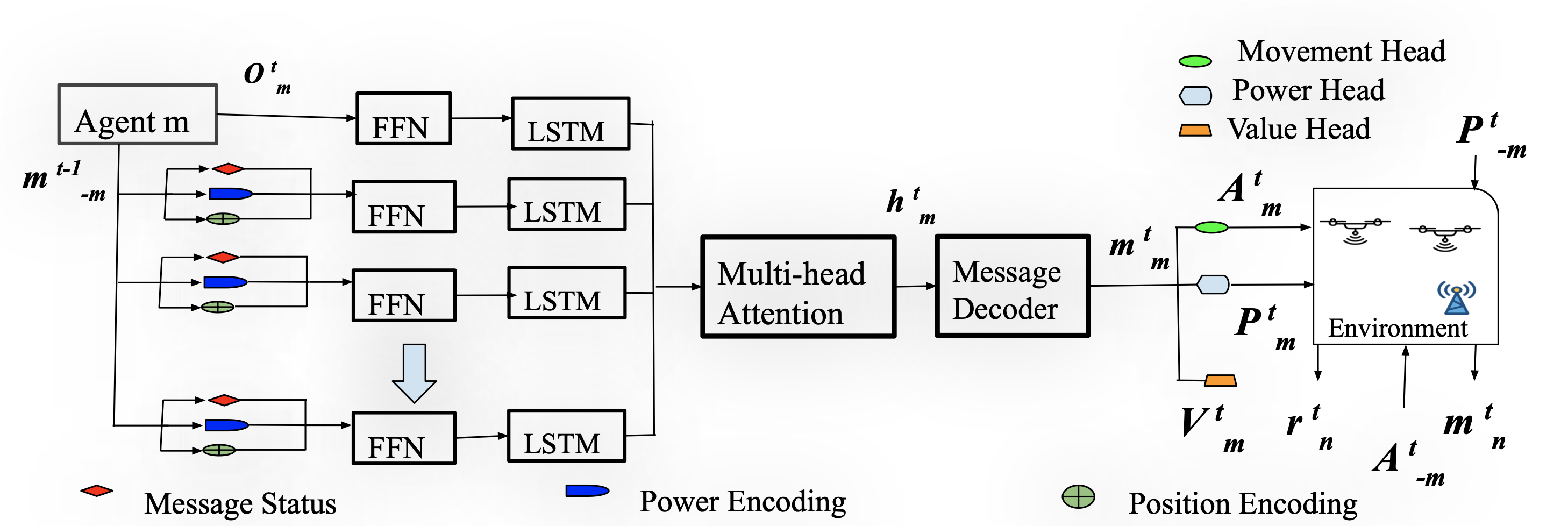} 
    \caption{Illustration of the architecture of our proposed \ac{MARL} algorithm with communication.}
    \label{fig:MARL_architecture}
\end{figure*}
For our \ac{MARL} framework, we employ a \ac{NN} architecture to learn communication-based action policies for each \ac{UAV} in the swarm. These policies are based on both the partial observations of the agents and the messages broadcasted.
At time step $t$, the $m^{th}$ agent receives its local observation $o_{m}^{t}$ along with messages from all other agents transmitted at time step $t-1$, denoted as $m_{\text{-}m}^{t-1}$. Here, the notation $-m$ represents the set of all agents except the $m^{th}$ agent.
The local observation of each agent comprises its current position (X and Y coordinates) and, for each of the $q_{\max}$ detected targets, their positions, their sensed \ac{SNR}, and the X and Y coordinates of the corresponding \acp{BS}. Formally, the local observations can be expressed as
\begin{align*}
 o_m^t &= p_{m}^t \cup \left( \bigcup\limits_{i=1}^{q_{\max}} \{ \pmb{\gamma}_{m,b_{m}}^{t}(i), p_{t_i}^{t}, p_{b_m}\} \right) \in \mathbf{R}^{2 + 6q_{\max}},
\label{eq:ag_observation}
\end{align*}
where $b_m$ is the index of the \ac{BS} with the link providing the highest \ac{SNR}  $\pmb{\gamma}_{m,b_m}^{t}(i)$ for target $i$. Additionally, $p_{m}^t$, $p_{t_i}^{t}$, and $p_{b_m}$ denote the positions of the $m^{th}$ \ac{UAV}, the $i^{th}$ target, and the corresponding \ac{BS}, respectively.
Furthermore, for each message transmitted by other agents, we account for the possibility that the $m^{th}$ agent may not receive the message. To address this, we construct adapted versions of the messages, where each adapted message consists of the received message or a randomly generated one if the original message cannot be decoded, along with the status of the message, and the index of the transmitting agent. Formally, for agent $j \in \text{-}m$, assuming the dimensionality of the transmitted messages is predefined as $s$, the adapted message $\Tilde{m}_{j}^{t} \in \mathbf{R}^{s+2}$ is constructed as follows
\begin{align*}
    \Tilde{m}_{j}^{t} = 
    \begin{cases}
        \left[m_{j}^{t-1}, 1, j \right] , & \text{if   } \pmb{\Gamma}_{j,m}^{t} \geq \pmb{\Gamma}_{2}, \\
        \left[\epsilon, 0, j \right], & \text{if   } \pmb{\Gamma}_{1} < \pmb{\Gamma}_{j,m}^{t} < \pmb{\Gamma}_{2}, \\
        \left[\epsilon, -1, j \right], & \text{if  } \pmb{\Gamma}_{j,m}^{t} < \pmb{\Gamma}_{1},
    \end{cases}
\end{align*}    
where $\epsilon \sim \mathcal{N}(0, \mathbf{I})$, and $\pmb{\Gamma}_{1}$ and $\pmb{\Gamma}_{2}$ are the communication \ac{SINR} thresholds as defined in Section~\ref{sec:sys_model}.
As depicted in Fig.~\ref{fig:MARL_architecture}, an \ac{LSTM}-Attention based architecture \cite{LSTM,Attention}, similar to \cite{Naoumi2023TANAGERS}, is used to enhance the learning of communication-based action policies for the \ac{UAV} swarm. This architecture outputs a distribution over actions, denoted as $\pi_{\left(\mathcal{A}^{m}, \mathcal{M}^{m}\right)}^{\Theta} \left( \cdot \mid o_{m}^{t} \right)$ for each agent $m$ at time step $t$, as well as the messages to be transmitted for the subsequent time step.
For the $m^{th}$ agent, the partial observation $o_{m}^{t}$ is initially encoded using an observation-specific module, which consists of a single-layer \ac{FFN} $\pmb{o}_{\text{FFN}}\left(.\right)$ followed by an \ac{LSTM} module $\pmb{o}_{\text{LSTM}}\left(.\right)$. This module maps the partial observation input to an embedding $\Tilde{o}_m^t$ of size $s$, corresponding to the predefined communication message size.
Simultaneously, the adapted messages from other agents $\Tilde{m}_{j}^{t}, \forall j \in \text{-}m$ are encoded using a message-specific module, which also includes a single-layer \ac{FFN} $\pmb{m}_{\text{FFN}}\left(.\right)$ and an \ac{LSTM} cell $\pmb{m}_{\text{LSTM}}\left(.\right)$, mapping the adapted messages to an embedding $\dbtilde{m}_{j}^{t}$ of size $s$.
Subsequently, a multi-head attention block with $M$+$1$ heads aggregates the observation representation with the encoded messages from other agents. The output of this attention block, denoted as $\dbtilde{h}_m^{t}$, is then fed into a message decoder, a two-layer \ac{FFN} with a ReLU activation function denoted as $\pmb{m}_{\text{dec}}\left(.\right)$. This decoder outputs the outgoing message $m_{m}^{t}$ for the next step. This message also serves as an input to both the policy and value heads.
Indeed, two policy heads are employed, each composed of an \ac{FFN} followed by a softmax function to output distributions over the action space. The first policy head, $\pi_{mov} \left(\cdot\right)$, is responsible for path planning, outputting a distribution over possible \ac{UAV} movement actions. In our work, we consider that \acp{UAV} operate at a fixed altitude and constant speed $\nu$. Thus, the discrete action space includes movements along the X and Y axes and diagonal movements represented as $\{\left(\pm \nu, 0\right), \left(0, \pm \nu\right), (\pm \frac{1}{\sqrt{2}}\nu, \pm \frac{1}{\sqrt{2}}\nu)\}$.
The second policy head $\pi_{pow} \left(\cdot\right)$, serves as a transmission power adaptation mechanism, outputting a distribution over the set of possible communication transmit power levels. At each time step, movement and power level actions $a_m^{t} = \left(A_m^{t}, P_m^{t}\right)$ are sampled for each agent from these policy distributions.
Additionally, the value head, denoted as $\mathcal{V}\left(\cdot\right)$, is an \ac{FFN} used to estimate the value function $V_m^{t}$ and serves as a baseline for the \ac{MARL} algorithm, thereby enhancing the robustness and efficiency of training.
A summary of the formalized operations of the architecture is given in \textbf{Algorithm \ref{alg:alg2}}.
\begin{algorithm}[!h]
\caption{Operations of the \ac{MARL} architecture for the $m^{th}$ \ac{UAV} at time step $t$.}\label{alg:alg2}
\begin{algorithmic}[1]
\STATE \textbf{Input:} $o_m^t$, $\{ m_j^{t\text{-}1} \}_{j \in \text{-}m}$
\STATE \textbf{Observation and Message encoding:}
\STATE \hspace{0.5cm} $\Tilde{o}_m^t, c_m^t=\pmb{o}_{\text{LSTM}}(\pmb{o}_{\text{FFN}}(o_m^t), \Tilde{o}_i^{t-1}, c_m^{t-1})$
\STATE \hspace{0.5cm} $\dbtilde{m}_{j}^{t}, \Tilde{c}_j^t=\pmb{m}_{\text{LSTM}}(\pmb{m}_{\text{FFN}}(\Tilde{m}_{j}^{t}), \dbtilde{m}_{j}^{t-1}, \Tilde{c}_j^{t-1}), \forall j \in \text{-}m$
\STATE \textbf{Multi-Head Attention:}
\STATE \hspace{0.5cm} $Q, K, V =  \Tilde{o}_m^t \cup (\mathop{\cup}_{j \in \text{-}m} \dbtilde{m}_{j}^{t} )$
\STATE \hspace{0.5cm} $\dbtilde{h}_m^{t} = \operatorname{MultiHead} (Q,K,V)$
\STATE \textbf{Message decoding:}
\STATE \hspace{0.5cm} $m_m^{t} = \pmb{m}_{\text{dec}} (\dbtilde{h}_m^{t})$
\STATE \textbf{Action selection:}
\STATE \hspace{0.5cm} $A_m^{t} \sim \pi_{mov} (m_m^{t}),   P_m^t \sim \pi_{pow}(m_m^{t})$
\STATE \textbf{Value computation:}
\STATE \hspace{0.5cm} $V_m^{t} = \mathcal{V}\left(m_m^{t}\right)$
\end{algorithmic}
\end{algorithm}
\noindent Here, $c_m^t$ and $\tilde{c}_j^t$ are the cell states of the \ac{LSTM} modules, and they are randomly initialized at $t=0$, along with the encoded representations $\tilde{o}_l^0$ and $\tilde{c}_l^0$ for all agents $l \leq M$.
The $\operatorname{MultiHead}$ attention operation is defined as 
$
\operatorname{MultiHead}(Q, K, V) = \operatorname{Concat}(\operatorname{head}_1, \ldots, \operatorname{head}_{M\operatorname{+}1})W^O
$
\noindent where 
\begin{equation*}
\begin{split}
  \operatorname{head}_i
    & = 
   \operatorname{Softmax}\left(\frac{QW_i^Q \left( KW_i^K \right)^T}{\sqrt{d_k}}\right) VW_i^V,
\end{split}
\end{equation*}
\noindent where $d_k$ is the dimension of the inputs and $W^O$, $W_i^Q$, $W_i^K$, and $W_i^V$ are parameter matrices to be learned.
For training, we use the policy gradient method to update the parameters $\Theta$ of the architecture, thereby optimizing the reward defined in Eq.~\eqref{eq:discounted_reward}. Specifically, the parameters of the architecture are updated by minimizing the following loss function
\small
\begin{equation*}
\begin{split}
& \nabla_{\Theta} \mathcal{L}(\Theta) 
=
\frac{1}{T_{\max }^\prime} \sum_{m=1}^M \sum_{t=1}^{T_{\max}^\prime}\Bigl[-\nabla_\theta \left(\log \pi_{mov}\left(A_m^t \mid o_m^t\right) \right. 
+ 
\\ & 
\left. \log \pi_{pow}\left(P_m^t \mid o_m^t\right) \right)
\times %
 \left(\pmb{\mathcal{R}}^{t}-V_m^t\right)+\beta \nabla_\Theta\left(\pmb{\mathcal{R}}^{t}-V_m^t\right)^2\Bigr] ,
\end{split}
\label{eq:loss_func}
\end{equation*}
\normalsize
where $\pmb{\mathcal{R}}^t$ is the total achieved \ac{SNR} reward, and $T_{\max}^\prime$ is the number of iterations within a batch. The policy gradient loss function combines both the movement action and transmission power policy losses, along with the value loss, balanced by the coefficient $\beta$. Additionally, the architecture parameters are shared across \acp{UAV} to improve the training efficiency.
\section{Results} 
\label{sec:numerical_results}
To demonstrate the effectiveness of our proposed approach, we evaluate it on a specific use case focused on monitoring drones within a defined environment. The parameters for this environment are provided in Table~\ref{table:simulation_parameters}.
Additionally, we consider various types of commercial drones and their corresponding \ac{RCS} distributions, based on the data from \cite{rcs_drones}.
It is worth noting that the results presented in this section are averaged across multiple independent simulation runs to ensure the robustness of the proposed approach.
\begin{table}[!t]
\centering
\caption{Simulation and Training Parameters.}
\begin{tabular}{@{}ll@{}}
\toprule
\textbf{Simulation Parameter}    & \textbf{Value}     \\ 
\midrule
Environment dimensions ($L \times H$) & $1000 \! \times \! 1000 \ \si{\metre \squared}$ \\
Number of iterations ($T_{\max}$)               & $100$            \\
UAVs altitude ($h$)               & $25 \ \si{\metre}$            \\
UAVs speed ($\nu$)                & $20 \ \si{\metre\per\second}$           \\
Carrier frequency ($f_c$)        &  $28 \ \si{\GHz}$             \\
Shadow fading standard deviation ($\sigma$)        &  $ 3.56 \ \si{\dB}$             \\
Transmit Power (${\mathcal{P}}_{tx}^{k}$)           &      $46 \ \si{\dB}m$     \\
Noise Power ($P_n$)              &     $-99$ \ \si{\dB}m        \\
Antenna gains ($\mathcal{G}_{tx}= \mathcal{G}_r$)   &     $11 \ \si{\dB}i$         \\
Communication \ac{SINR} thresholds ($\pmb{\Gamma}_1 < \pmb{\Gamma}_2$)   &   $(-10, 0) \  \si{\dB}$  \\
Sensing \ac{SNR} threshold ($\pmb{\gamma}_{s}$)   &   $-10 \ \si{\dB}$  \\
\midrule \midrule
\textbf{Training Parameter}      & \textbf{Value}     \\ 
\midrule
Discount factor ($\gamma$)              &  $0.9$                \\
Optimizer              &  $\operatorname{RMSProp}$               \\
Number of epochs                       &    $100$                \\
Batch size                       &    $5$                        \\
Learning rate                    &  $10^{-4}$                    \\ 
Value loss coefficient ($\beta$)                    &  $0.014$   \\ 
\bottomrule
\end{tabular}
\label{table:simulation_parameters}
\end{table}

\begin{figure}[!t]
\centering
\includegraphics[scale=0.32]{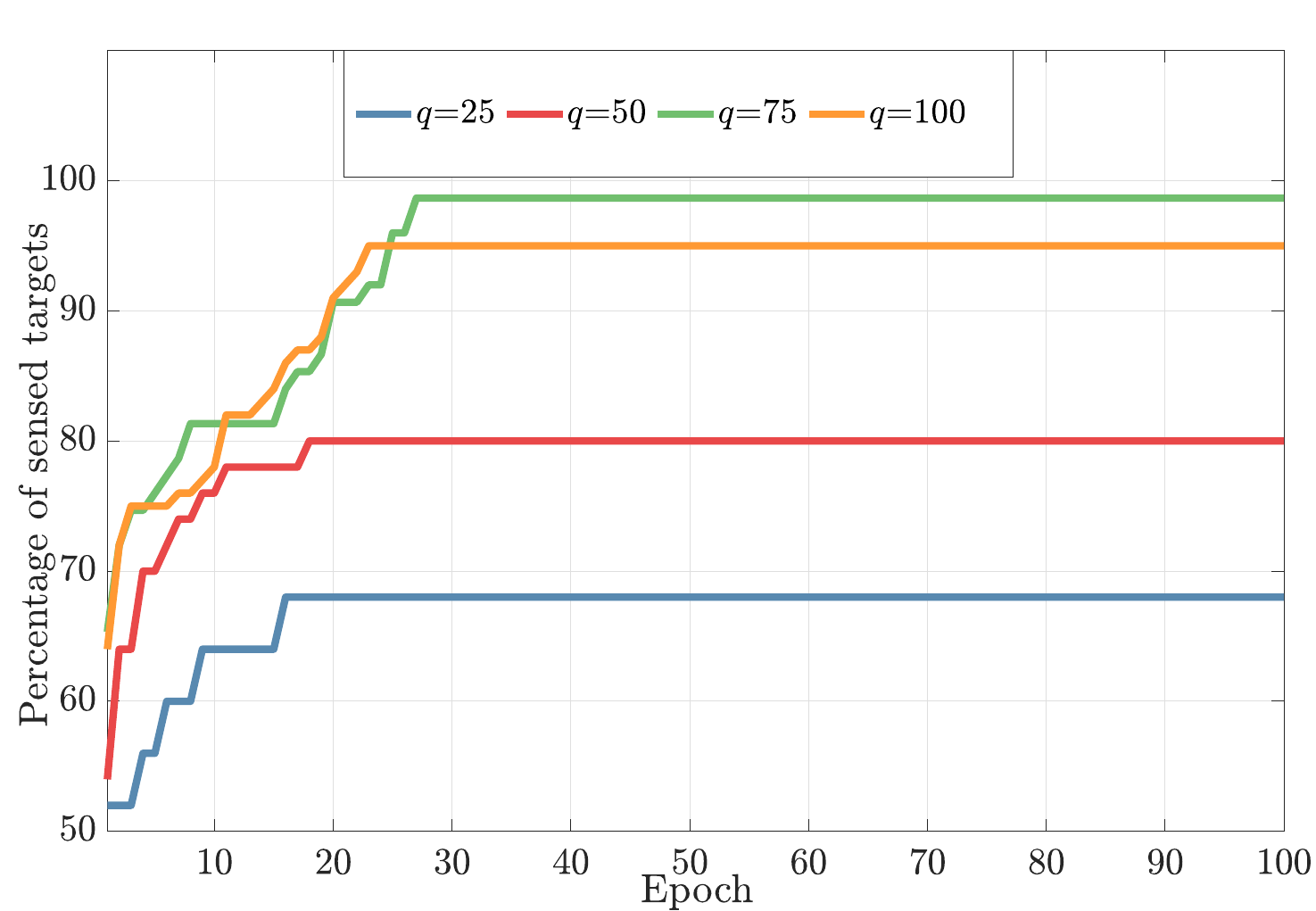}
\caption{Evolution of the percentage of detected targets by the \ac{UAV} swarm in environments with the number of targets $q$.}
\label{fig:percentage_targets}
\end{figure}

First, we examine the sensing efficiency of \acp{UAV} in environments with varying numbers of targets. The results shown in Fig.~\ref{fig:percentage_targets} illustrate the convergence of the \ac{MARL} algorithm in terms of the percentage of covered targets out of a varying total number $q$ over training epochs. Remarkably, the algorithm achieves over $70$\% target coverage in most experiments within a short number of epochs, demonstrating both efficient training and fast convergence. For instance, it requires fewer than $30$ epochs to achieve nearly $100$\% coverage in an environment with $75$ targets. In these experiments, the number of deployed \acp{UAV} is fixed at $5$. The observed performance differences in the final percentage of covered targets are mainly attributed to the spatial distribution of targets within the environment and the constraints imposed by the maximum flight duration, which limit the ability of the \ac{UAV} swarm to thoroughly explore the environment. Nevertheless, a higher number of targets allows for greater coverage during the planned trajectories of the \ac{UAV} swarm.
Furthermore, we analyze the discounted cumulative reward of all \acp{UAV} in environments with varying numbers of \acp{UAV} $M$. In this analysis, the number of targets is set to $100$, and the resulting reward is plotted over $100$ epochs. The results show that increasing the number of agents ensures that more targets are covered, yielding higher rewards compared to deploying fewer agents. This performance increase is primarily due to the efficient inter-communication incorporated into the \ac{MARL} algorithm, which maintains coordination among the agents and avoids scalability issues. The trade-off is that the algorithm requires more time to converge with a higher number of agents. However, the algorithm typically needs less than $30$ epochs to converge.
\begin{figure}[!t]
\centering
\includegraphics[scale=0.32]{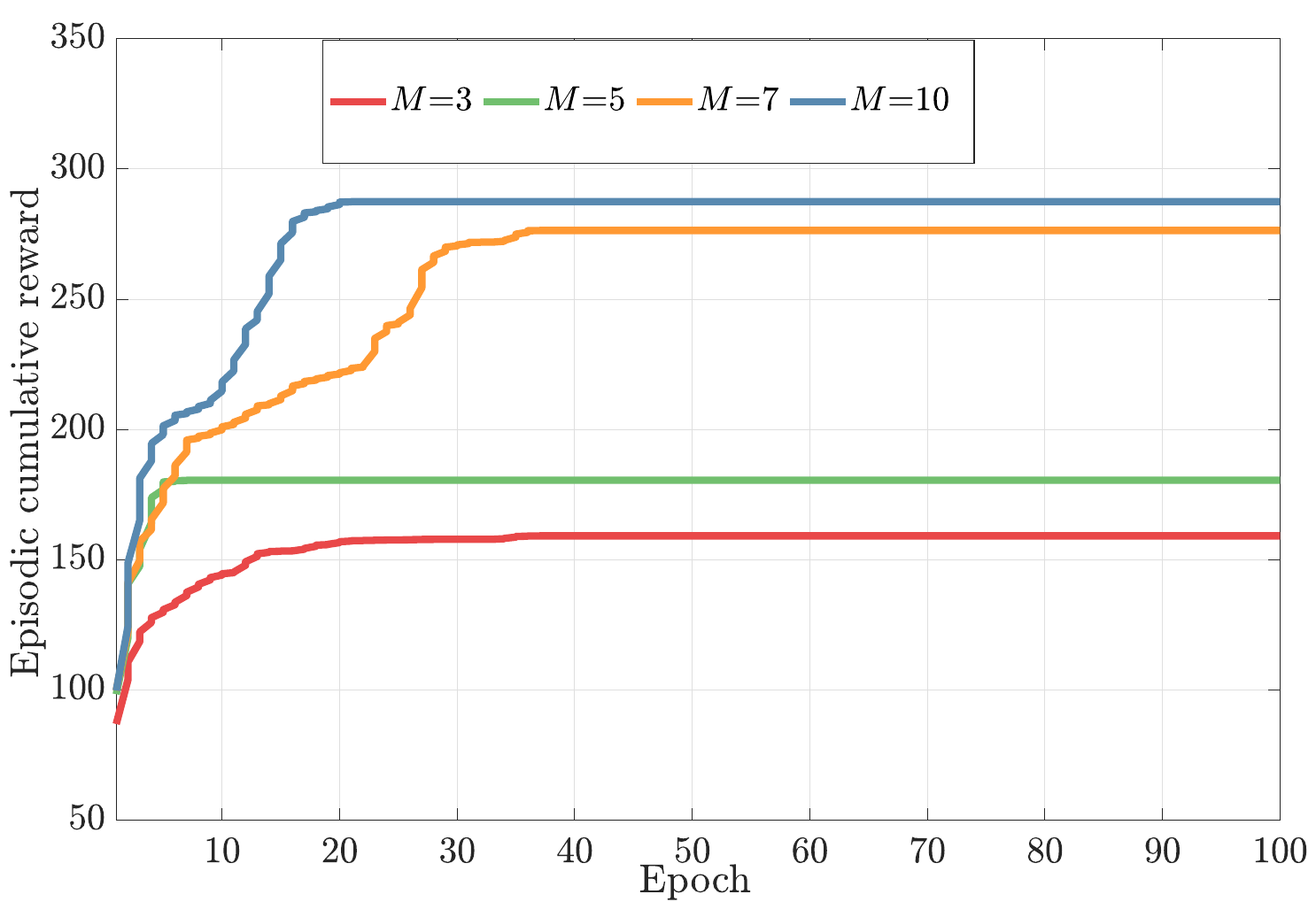}
\caption{Episodic cumulative reward comparison in environments with increasing \ac{UAV} number ($M$).}
\label{fig:cumulative_snr}
\end{figure}

\begin{figure}[!t]
\centering
\includegraphics[width=\linewidth]{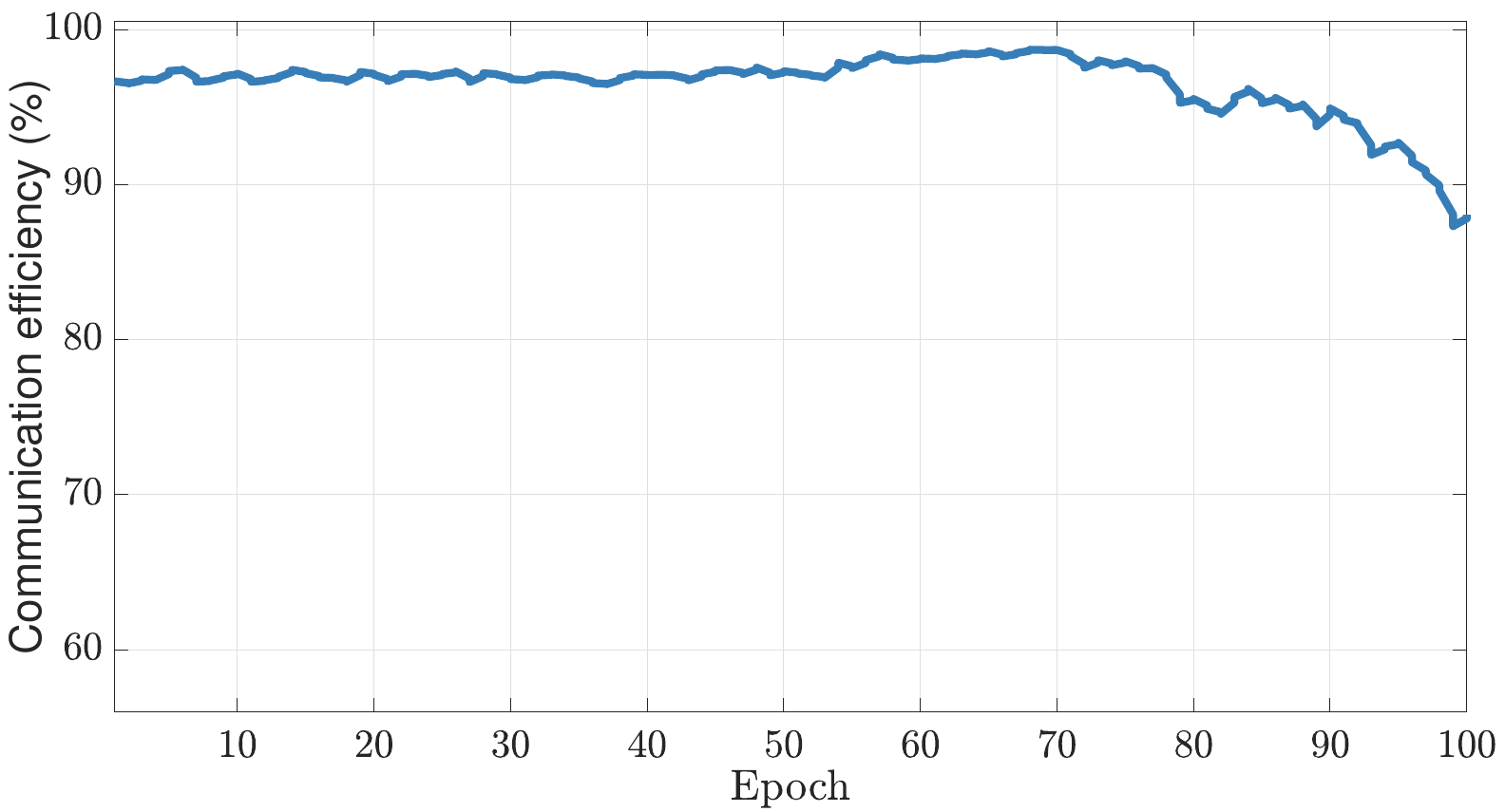}
\caption{Inter-\ac{UAV} communication efficiency as the percentage of messages exceeding the \ac{SINR} threshold.}
\label{fig:comm_efficiency}
\end{figure}

Finally, we evaluate the algorithm's ability to learn optimal power levels for each \ac{UAV} to maintain reliable communication by ensuring transmitted messages exceed the \ac{SINR} threshold required for decoding. 
Fig.~\ref{fig:comm_efficiency} shows the percentage of messages surpassing the \ac{SINR} threshold $\pmb{\Gamma}_2$, indicating successful transmission.
The algorithm maintains over $95\%$ efficiency up to the $80$th epoch, before dropping to a minimum of $87\%$  by the 100th epoch. This decline occurs when all targets in the environment have been sensed, reducing the need for active \ac{UAV} communication and movement. Despite this, the algorithm remains robust, effectively balancing power usage based on operational needs. These findings highlight the robust performance of the proposed framework, demonstrating its suitability for \ac{ISAC} applications and confirming its effectiveness and broad applicability.
\section{Conclusion} \label{sec:conclusions}
This paper introduced a novel \ac{MARL} framework for \ac{UAV} swarm path planning in communication-centric bistatic \ac{ISAC} applications. Our approach leverages inter-\ac{UAV} communications to optimize positioning and path planning, thereby enhancing the overall radar sensing performance. The decentralized cooperative framework enables \acp{UAV} to develop effective communication protocols, improving their environmental awareness and operational efficiency. Notably, our framework demonstrates exceptional performance in terms of coverage and overall sensing \ac{SNR} across various environmental settings, as well as resilience to realistic wireless conditions by adapting to environmental changes and unstable communication links. Additionally, a transmission power adaptation technique ensures robust performance and maximizes \ac{SINR} for \ac{UAV} communications. These contributions underscore the framework's robustness, efficiency, and broad applicability for \ac{ISAC} networks.


\begin{thebibliography}{10}
\providecommand{\url}[1]{#1}
\csname url@samestyle\endcsname
\providecommand{\newblock}{\relax}
\providecommand{\bibinfo}[2]{#2}
\providecommand{\BIBentrySTDinterwordspacing}{\spaceskip=0pt\relax}
\providecommand{\BIBentryALTinterwordstretchfactor}{4}
\providecommand{\BIBentryALTinterwordspacing}{\spaceskip=\fontdimen2\font plus
\BIBentryALTinterwordstretchfactor\fontdimen3\font minus \fontdimen4\font\relax}
\providecommand{\BIBforeignlanguage}[2]{{%
\expandafter\ifx\csname l@#1\endcsname\relax
\typeout{** WARNING: IEEEtran.bst: No hyphenation pattern has been}%
\typeout{** loaded for the language `#1'. Using the pattern for}%
\typeout{** the default language instead.}%
\else
\language=\csname l@#1\endcsname
\fi
#2}}
\providecommand{\BIBdecl}{\relax}
\BIBdecl

\bibitem{LiuJSAC}
F.~Liu, Y.~Cui, C.~Masouros, J.~Xu, T.~X. Han, Y.~C. Eldar, and S.~Buzzi, ``Integrated sensing and communications: Toward dual-functional wireless networks for 6g and beyond,'' \emph{IEEE Journal on Selected Areas in Communications}, vol.~40, no.~6, pp. 1728--1767, 2022.

\bibitem{zhang2022accelerating}
T.~Zhang, S.~Wang, G.~Li, F.~Liu, G.~Zhu, and R.~Wang, ``{Accelerating Edge Intelligence via Integrated Sensing and Communication},'' 2022.

\bibitem{jump}
J.~Pegoraro \emph{et~al.}, ``{JUMP: Joint Communication and Sensing With Unsynchronized Transceivers Made Practical},'' \emph{IEEE Transactions on Wireless Communications}, vol.~23, no.~8, pp. 9759--9775, 2024.

\bibitem{surveyUAVs}
X.~Gu and G.~Zhang, ``{A survey on UAV-assisted wireless communications: Recent advances and future trends},'' \emph{Comput. Commun.}, vol. 208, no.~C, p. 44–78, aug 2023.

\bibitem{JAndrewZhang2020EnablingJC}
J.~A. Zhang, M.~L. Rahman, K.~Wu, X.~Huang, Y.~J. Guo, S.~Chen, and J.~Yuan, ``{Enabling Joint Communication and Radar Sensing in Mobile Networks—A Survey},'' \emph{IEEE Communications Surveys \& Tutorials}, vol.~24, no.~1, pp. 306--345, 2022.

\bibitem{uav_bs1}
H.~Shakhatreh, A.~Khreishah, and B.~Ji, ``{Providing wireless coverage to high-rise buildings using UAVs},'' in \emph{2017 IEEE International Conference on Communications (ICC)}, 2017, pp. 1--6.

\bibitem{yang2022path}
S.~Yang, Z.~Shan, J.~Cao, Y.~Gao, Y.~Guo, P.~Wang, X.~Wang, J.~Wang, T.~Zhang, and J.~Guo, ``{Path planning of UAV base station based on deep reinforcement learning},'' \emph{Procedia Computer Science}, vol. 202, pp. 89--104, 2022.

\bibitem{isac_uav1}
Y.~Jiang, Q.~Wu, W.~Chen, and K.~Meng, ``{UAV-Enabled Integrated Sensing and Communication: Tracking Design and Optimization},'' \emph{IEEE Communications Letters}, vol.~28, no.~5, pp. 1024--1028, 2024.

\bibitem{Naoumi2023TANAGERS}
S.~Naoumi, R.~Bomfin, R.~Alami, and M.~Chafii, ``{TANAGERS: Emergent Communication for UAVs as Flying Passive Radars},'' in \emph{IEEE WCNC Model-driven DL for 6G IoT}, 2023.

\bibitem{isac_uav2}
J.~Mu, R.~Zhang, Y.~Cui, N.~Gao, and X.~Jing, ``Uav meets integrated sensing and communication: Challenges and future directions,'' \emph{IEEE Communications Magazine}, vol.~61, no.~5, pp. 62--67, 2023.

\bibitem{chafii2023emergent}
M.~Chafii, S.~Naoumi, R.~Alami, E.~Almazrouei, M.~Bennis, and m.~Debbah, ``Emergent communication in multi-agent reinforcement learning for future wireless networks,'' \emph{IEEE Internet of Things Magazine}, vol.~6, pp. 18--24, 12 2023.

\bibitem{comm_pl}
M.~Polese, L.~Bertizzolo, L.~Bonati, A.~Gosain, and T.~Melodia, ``An experimental mmwave channel model for uav-to-uav communications,'' \emph{CoRR}, vol. abs/2007.11869, 2020.

\bibitem{SINRpaper}
Y.~Cohen, T.~Gafni, R.~Greenberg, and K.~Cohen, ``{SINR-Aware Deep Reinforcement Learning for Distributed Dynamic Channel Allocation in Cognitive Interference Networks},'' \emph{ArXiv}, vol. abs/2402.17773, 2024.

\bibitem{LSTM}
S.~Hochreiter and J.~Schmidhuber, ``Long short-term memory,'' \emph{Neural computation}, vol.~9, no.~8, pp. 1735--1780, 1997.

\bibitem{Attention}
A.~Vaswani, N.~M. Shazeer, N.~Parmar, J.~Uszkoreit, L.~Jones, A.~N. Gomez, L.~Kaiser, and I.~Polosukhin, ``{Attention is All you Need},'' in \emph{Neural Information Processing Systems}, 2017.

\bibitem{rcs_drones}
V.~Semkin, J.~Haarla, T.~Pairon, C.~Slezak, S.~Rangan, V.~Viikari, and C.~Oestges, ``Analyzing radar cross section signatures of diverse drone models at mmwave frequencies,'' \emph{IEEE Access}, vol.~8, pp. 48\,958--48\,969, 2020.

\end{thebibliography}


\end{document}